\documentclass[a4paper,12pt]{article}
\usepackage[pctex32]{graphics}
\usepackage{amssymb,amsmath}

\def\ben{\begin{equation}}
\def\een{\end{equation}}

\def\nn{\nonumber} \def\bd{\begin{document}} \def\ed{\end{document}}
\def\ds{\documentstyle} \let\fr=\frac \let\bl=\bigl \let\br=\bigr
\let\Br=\Bigr \let\Bl=\Bigl
\let\bm=\bibitem
\let\na=\nabla
\let\pa=\partial \let\ov=\overline
\newcommand{\be}{\begin{equation}}
\newcommand{\ee}{\end{equation}}
\def\ba{\begin{array}}
\def\ea{\end{array}}
\def\ft#1#2{{\textstyle{\frac{\scriptstyle #1}{\scriptstyle #2} } }}
\def\fft#1#2{{\frac{#1}{#2}}}
\def\del{\partial}
\def\vp{\varphi}
\def\sst#1{{\scriptscriptstyle #1}}
\def\oneone{\rlap 1\mkern4mu{\rm l}}
\def\td{\tilde}
\def\wtd{\widetilde}
\def\ie{{\it i.e.\ }}
\def\dalemb#1#2{{\vbox{\hrule height .#2pt
        \hbox{\vrule width.#2pt height#1pt \kern#1pt
                \vrule width.#2pt}
        \hrule height.#2pt}}}
\def\square{\mathord{\dalemb{6.8}{7}\hbox{\hskip1pt}}}
\newcommand{\ho}[1]{$\, ^{#1}$}
\newcommand{\hoch}[1]{$\, ^{#1}$}
\newcommand{\bea}{\setlength\arraycolsep{2pt} \begin{eqnarray}}
\newcommand{\eea}{\end{eqnarray}}
\newcommand{\ra}{\rightarrow}
\newcommand{\lra}{\longrightarrow}
\newcommand{\Lra}{\Leftrightarrow}
\newcommand{\bp}{\tilde \beta^\prime}
\newcommand{\tr}{{\rm tr} }
\newcommand{\Tr}{{\rm Tr} }
\def\0{{\sst{(0)}}}
\def\1{{\sst{(1)}}}
\def\2{{\sst{(2)}}}
\def\3{{\sst{(3)}}}
\def\4{{\sst{(4)}}}
\def\5{{\sst{(5)}}}
\def\6{{\sst{(6)}}}
\def\7{{\sst{(7)}}}
\def\8{{\sst{(8)}}}
\def\m{{\sst{(m)}}}
\def\n{{\sst{(n)}}}
\def\cA{{{\cal A}}}
\def\cB{{{\cal B}}}
\def\cF{{{\cal F}}}
\def\cG{{{\cal G}}}
\def\cH{{{\cal H}}}
\def\tV{\widetilde V}
\def\tW{\widetilde W}
\def\tH{\widetilde H}
\def\tE{\widetilde E}
\def\tF{\widetilde F}
\def\tA{\widetilde A}
\def\im{{{\rm i}}}
\def\tY{{{\wtd Y}}}
\def\ep{{\epsilon}}
\def\vep{{\varepsilon}}
\def\bD{{{\bar D}}}
\def\R{{{\mathbb R}}}
\def\C{{{\mathbb C}}}
\def\H{{{\mathbb H}}}
\def\CP{{{\mathbb C}{\mathbb P}}}
\def\RP{{{\mathbb R}{\mathbb P}}}
\def\Z{{{\mathbb Z}}}
\def\bA{{{\mathbb A}}}
\def\bB{{{\mathbb B}}}
\def\bC{{{\mathbb C}}}
\def\bD{{{\mathbb D}}}
\def\bE{{{\mathbb E}}}
\def\bZ{{{\mathbb Z}}}
\def\Re{{{\frak{Re}}}}
\def\Im{{{\frak{Im}}}}
\def\cosec{{\,\hbox{cosec}\,}}
\def\Gm{{\Gamma_{\!\! -}}}
\def\Gp{{\Gamma_{\!\! +}}}
\def\stan{{standard }}
\def\nonstan{{supernumerary }}
\def\p{{\partial}}
\def\kdel#1{{\fft{\del}{\del#1}}}

\def\bog{{Bogomolny }}
\def\om{{\omega}}

\newcommand{\nnr}{\nonumber \\}
\newcommand{\pd}{\partial}
\newcommand{\ud}{\textrm{d}}
\newcommand{\dTH}{T^{\prime \, 0}_\textrm{H}}
\newcommand{\dOi}{\Omega^{\prime \, 0}_i}
\newcommand{\bx}{{\bf x}}

\begin{document}
\begin{titlepage}
\vspace{5mm}
\begin{center}
{\Large \bf Slowly rotating black holes \\
 in the Ho\v{r}ava-Lifshitz gravity }

\vskip .6cm
 \centerline{\large
 Hyung Won Lee,  Yong-Wan Kim, Yun Soo Myung}

\vskip .6cm

{Institute of Basic Science and School of Computer Aided Science,
\\Inje University, Gimhae 621-749, Korea \\}

\end{center}

\vspace{5mm}

\begin{abstract}
We investigate  slowly rotating black holes in the
Ho\v{r}ava-Lifshitz (HL) gravity.  For $\Lambda_W=0$ and
$\lambda=1$, we find a slowly rotating black hole  of the
Kehagias-Sfetsos solution in asymptotically flat spacetimes.  We
discuss their thermodynamic properties by computing mass,
temperature,  angular momentum, and angular velocity on the horizon.
\end{abstract}
\end{titlepage}

\section{Introduction}
Ho\v{r}ava has proposed a renormalizable theory of gravity at a
Lifshitz point~\cite{ho1,ho2},  which  may be regarded as a UV
complete candidate for general relativity. At short distances the
theory of  Ho\v{r}ava-Lifshitz (HL) gravity describes interacting
nonrelativistic gravitons and is supposed to be power counting
renormalizable in (1+3) dimensions. Recently, its black hole
solutions  has been intensively investigated in
~\cite{LMP,CCO1,CLS,CY,MK,KS,CCO2,Gho,Myung,CJ1,CJ2,CW,park,BGS,Gho09,CL,PW,lkme,Myungent,CK}.

Considering spherically symmetric spacetimes, L\"u-Mei-Pope (LMP)
have obtained the black hole solution with dynamical parameter
$\lambda$~\cite{LMP} and topological black holes were found in
\cite{CCO1}. Its thermodynamics were studied in \cite{CCO2}, but
there remain unclear issues in defining the ADM mass and entropy
because their asymptotes are Lifshitz~\cite{MK,lifMyung}. On the
other hand, Kehagias and Sfetsos (KS) have found the $\lambda=1$
black hole solution in asymptotically flat spacetimes using the
deformed HL gravity~\cite{KS}. Its thermodynamics was discussed in
Ref.\cite{Myung,Myungent}. On later, Park has obtained a $\lambda=1$
black hole solution with two parameter $\omega$ and
$\Lambda_W$~\cite{park} and the authors in~\cite{CK} have found that
the BTZ black strings are solutions to the HL gravity. The most
general spherically symmetric solution with zero shift vector  was
found in the non-projectable Ho\v{r}ava-Lifshitz class of theories
with general coupling constants for the quadratic terms~\cite{KKBH}.

 It is very curious to find
a rotating black hole solution in the HL gravity. However, it seems
to be a formidable task to find a fully rotating solution because
equations of motion to be solved are very complicated. In this work,
we wish to find a slowly rotating black hole solutions based on the
KS solutions  by introducing a non-zero shift vector. Here ``slowly
rotating" black hole means that we consider up to linear order of
rotating parameter $a=J/M(a\ll1)$ in the metric functions, equations
of motion, and thermodynamic
quantities~\cite{HH,Pois,Mart,GS,KC,slMyung}. We mention that the
slowly rotating Kerr
 black hole is recovered from the slowly
rotating black hole solutions in the HL gravity, in the IR limit of
$\omega \to \infty(\kappa^2 \to 0)$.

\section{HL gravity}
In this section, we review briefly the HL gravity including the soft
violation term. In the ADM formalism, the four dimensional metric of
general relativity is parameterized as \cite{adm}
\be
ds_4^2= - N^2  dt^2 + g_{ij} (dx^i - N^i dt) (dx^j - N^j dt)\,,
\label{metricans}
\ee
where the lapse, shift and 3-metric $N$, $N^i$, and $g_{ij}$ are all
functions of $t$ and $x^i$.  In the simplest version of the theory
which respects the projectability condition~\cite{ho1,ho2}, the
lapse function $N$ is viewed as a gauge field for time
re-parameterizations, and it is effectively restricted to depend
only on $t$.  A closer parallel with general relativity could be
achieved if this projectability restriction is relaxed. Thus one may
take a broader view of the Ho\v{r}ava proposal without this
condition as an interesting  class of theories. Then,   the relative
coefficients of the terms in the ADM decomposition of the
Einstein-Hilbert action are modified, and additional terms involving
higher spatial derivatives are included too.  The higher derivative
terms may improve the renormalizability of the theory without the
problems of ghosts that would arise if higher temporal derivatives
were present.

   The ADM decomposition  of the
Einstein-Hilbert action is given by
\be
S_{EH} = \fft{1}{16\pi G} \int d^4x \sqrt{g} N (K_{ij} K^{ij} - K^2 + R -
2\Lambda)\,,\label{ehlag}
\ee
where $G$ is Newton's constant and extrinsic curvature $K_{ij}$ is
defined by
\be
K_{ij} = \fft{1}{2N} (\dot g_{ij} - \nabla_i N_j - \nabla_j N_i)\,.
\ee
Here, a dot denotes a derivative with respect to $t$.

   The modified action of the
theory  including  $\mu^4 R$ can be written as
\bea%
\label{action} S&=&\int dtd^3\bx\, ({\cal L}_0 + {\cal L}_1)\,,\\
{\cal L}_0 &=& \sqrt{g}N\left\{\frac{2}{\kappa^2}(K_{ij}K^{ij}
-\lambda K^2)+\frac{\kappa^2\mu^2(\Lambda_W R
  -3\Lambda_W^2)}{8(1-3\lambda)} + \frac{\kappa^2\mu^2\omega R}{8(3\lambda-1)}\right\}\,,\nn\\ {\cal L}_1&=&
\sqrt{g}N\left\{\frac{\kappa^2\mu^2 (1-4\lambda)}{32(1-3\lambda)}R^2
-\frac{\kappa^2}{2W^4} \left(C_{ij} -\frac{\mu W^2}{2}R_{ij}\right)
\left(C^{ij} -\frac{\mu W^2}{2}R^{ij}\right)\right\}\,,\nn
\eea%
where $\lambda\,,\kappa\,,\mu\,,$ and $W$  are constant parameters
to denote the HL gravity and  $\Lambda_W$ is a negative cosmological
constant.   Here $\omega = 8(3\lambda-1)\mu^2/\kappa^2$ is a
parameter to represent a soft violation term of  the detailed
balance condition, $\mu^4 R$ and $C_{ij}$ is the Cotton tensor
defined by
\be
C^{ij}=\epsilon^{ik\ell}\nabla_k\left(R^j{}_\ell
-\frac14R\delta_\ell^j\right) \,.\label{def.K.C}
\ee
Two cases of  $\Lambda_W=0$ with $\lambda=1$ and $\omega=0$ are
included: the former case provides  the KS solution, while the
latter shows the LMP solution.  For the case of $\omega=0$,
comparing ${\cal L}_0$ to that of general relativity plus negative
cosmological constant in the ADM formalism, the speed of light,
Newton's constant, and the cosmological constant emerge
as~\cite{ho1}
\be c=\fft{\kappa^2\mu}{4}
\sqrt{\fft{\Lambda_W}{1-3\lambda}}\,,\qquad
G=\fft{\kappa^2}{32\pi\,c}\,,\qquad \Lambda=\ft32
\Lambda_W\,\label{cg} \ee which shows asymptotically AdS spacetimes.
On the other hand, for $\Lambda_W=0$ with $\lambda=1$, we have
asymptotically flat spacetimes as~\cite{KS} \be
c^2=\fft{\kappa^2\mu^4}{2}\,,\qquad
G=\fft{\kappa^2}{32\pi\,c}\,,\qquad \Lambda=0\,.\label{mcg} \ee

Since we wish to find a non-spherically symmetric solution of black
hole, it needs to  derive full equations of motion from the action
(\ref{action})~\cite{LMP}.  The equations of motion were also
obtained in~\cite{KK}. In deriving full equations, we have relaxed
both the projectability restriction and detailed balance condition
since
 the lapse function $N$  depends on the spatial
coordinates $x^i$ as well as a soft violation term of $\mu^4 R$ is
included.

\section{Slowly rotating black hole}
In this section, we  find  a slowly rotating black hole in
asymptotically flat spacetimes. Before we proceed, we would like to
mention a static solution, the KS solution with $\Lambda_W = 0$ and
$\lambda=1$.  In this case,  equation of motion  for the lapse
function $N$ can be read as
\be \fft{2}{\kappa^2}(K_{ij}K^{ij} - K^2) +\mu^4  R
+\frac{3\kappa^2\mu^2 }{64}R^2 - \frac{\kappa^2}{2W^4} Z_{ij}
Z^{ij}=0\,\label{eom1-KS}
\ee%
with \begin{equation} Z_{ij}=C_{ij} -\frac{\mu
W^2}{2}R_{ij}.\end{equation}
 The variation with respect to $\delta N^i$ implies
\be \nabla_k(K^{k\ell}-\,Kg^{k\ell})=0\,.\label{eom2-KS}
\ee%
The equations of motion  from  variation of $\delta g^{ij}$ takes
the form
\be%
\frac{2}{\kappa^2}E_{ij}^\1-\frac{2}{\kappa^2}E_{ij}^\2 +\mu^4
E_{ij}^\3 +\frac{3\kappa^2\mu^2}{64}E_{ij}^\4
-\frac{\mu\kappa^2}{4W^2}E_{ij}^\5
-\frac{\kappa^2}{2W^4}E_{ij}^\6=0\,,\label{eom3-KS}\ee%
where all $E_{ij}^{(r)}$ for $r=1,\cdots,6$ are the same as in
Ref.\cite{LMP} except the case  
\be E_{ij}^\3=N\Big(R_{ij}- \frac12Rg_{ij}\Big)-(
\nabla_i\nabla_j-g_{ij}\nabla_k\nabla^k)N\,.\ee
A spherically symmetric solution to these equations was obtained
by considering the line element
\begin{equation}
ds^2 = -N(r)^2 dt^2 + \frac{dr^2}{f(r)} + r^2 \left ( d \theta^2 + \sin^2 \theta d \phi^2 \right ) .
\label{sph_ansatz}
\end{equation}
In this case, we have $K_{ij}=0$ and $C_{ij}=0$.  The KS  solution
is given by
\begin{equation}
f_{\rm KS} =N^2_{\rm KS}= 1 + \omega r^2 \left ( 1 - \sqrt{1 +
\frac{4 M}{\omega r^3}}\right ) \label{sph_sol}
\end{equation}
with $\omega = 16 \mu^2 / \kappa^2$ and the mass parameter $M$. In
the limit of $\omega \to \infty$ (equivalently, the IR limit of
$\kappa^2 \to 0$), it reduces to the Schwarzschild form of \be
 f_{\rm Sch}(r)=1-\frac{2M_{\rm Sch}}{r}. \ee

Now let us introduce  an axisymmetric metric ansatz with one
component shift vector (or shift function) $N^\phi(r)$ as
\begin{equation}
ds^2 = -f(r) dt^2 + \frac{dr^2}{f(r)} + r^2 d \theta^2 +
       r^2 \sin^2 \theta \Big[ d \phi - a N^\phi(r) dt \Big]^2
\label{axi_ansatz}
\end{equation}
with the rotation parameter $a=J/M$.  Note that even the above
metric contains up to the second order of $a$, we will keep
equations of motion up to the linear order of $a$ in order to obtain
a slowly rotating black hole solution. This is equivalent to
considering the slowly rotating metric initially
\begin{equation}
ds^2_{\rm slow~R} = -f(r) dt^2 + \frac{dr^2}{f(r)} + r^2 d \theta^2
+
       r^2 \sin^2 \theta \Big[ d \phi^2 - 2a N^\phi(r) dt d\phi \Big].
\label{axisr_ansatz}
\end{equation}
For an axisymmetric metric ansatz, we cannot use the reduced
Lagrangian approach which was developed in~\cite{LMP} to find $f$
and $N$ because (\ref{axisr_ansatz})  contains a non-singlet
function $N^\phi$ under the SO(3) action on the $S^2$.  Hence,  we
have to solve Eqs.(\ref{eom1-KS}), (\ref{eom2-KS}), and
(\ref{eom3-KS}), simultaneously.

Using Eq.(\ref{axi_ansatz}), non-zero components of extrinsic
curvature are found to be
\begin{equation}
K_{ij} = \begin{pmatrix}
         0 & 0 & -\frac{1}{2} \frac{r^2 a \sin^2 \theta}{\sqrt{f(r)}} \frac{dN^\phi(r)}{dr} \\
         0 & 0 &0 & \\
         -\frac{1}{2} \frac{r^2 a \sin^2 \theta}{\sqrt{f(r)}} \frac{dN^\phi(r)}{dr} & 0 & 0
        \end{pmatrix},
\label{K_ij}
\end{equation}
where we observe that $K_{r\phi}=K_{\phi r}$ are  linear order of
$a$ and its trace is zero ($K=0$). Also we show that all components
of Cotton tensor still vanish ($C_{ij}=0$) up to linear order of
$a$. This may imply that if one wishes to find a fully rotating
black hole, all higher order terms of $a$ must be included.  Then,
Eq.(\ref{eom2-KS}) reduces to \be \nabla_k K^{k\ell}={\rm diag}
\left [ 0, 0, \frac{a \sqrt{f(r)}}{2r}
      \left (  r \frac{d^2 N^\phi(r)}{dr^2} + 4 \frac{dN^\phi(r)}{dr} \right ) \right ] = 0\,,\label{eom2_KS_reduce}
\ee%
which has a solution with two arbitrary constants $C_1$ and $C_2$
\begin{equation}
N^\phi(r) = C_1 + \frac{C_2}{r^3}.
\label{N3_sol}
\end{equation}
For later convenience, we choose the shift function to be \be
N^\phi(r) = \frac{2M}{r^3} \ee with $C_1=0$ and $C_2=2M$. In this
case, one has non-zero $g_{t\phi}$ and $g_{\phi t}$ components
\begin{equation}
g_{t\phi}=g_{\phi t}=-ar^2
N^\phi(r)\sin^2\theta=-\frac{2J}{r}\sin^2\theta.
\end{equation}
Plugging these into Eq.(\ref{eom1-KS}) leads to
\begin{equation}
\frac{(f-1)^2}{r^2}-\frac{2(f-1)f'}{r}-2\omega(1-f-rf') =
\frac{32a^2 M^2 \sin^2 \theta}{\kappa^4\mu^4r^4}.
\label{eom1_KS_reduce}
\end{equation}
We note that the right-hand side is taken  to be zero effectively
because it is second order of $a$.  Then, the solution is given by
\begin{equation}
f_\pm(r) = 1 + \omega r^2 \left ( 1  \pm
       \sqrt{1+\frac{C_3 }{\omega r^3}} \right ).
\label{sol_f11_KS}
\end{equation}
A metric function $f_-(r)$  recovers  the Schwarzschild black hole
solution in the limit of $\omega \to \infty$.  The
$(r,r)$-component of Eq.(\ref{eom3-KS}) gives the same equation as
in (\ref{eom1_KS_reduce}). Other two components of
$(\theta,\theta)$ and $(\phi, \phi)$ provide  second order
differential equations for the metric function $f(r)$ and solving
these leads to  the solution up to linear order $a$ as
\begin{equation}
f_\pm(r) = 1 + \omega r^2 \left ( 1  \pm
       \sqrt{1+\frac{2}{\omega r^2}\left ( 1-\frac{C_4}{16\mu^2} \right )
              +\frac{C_5 }{\omega r^3}} \right )
\label{sol_f22_KS}
\end{equation}
which becomes the same function as in (\ref{sol_f11_KS}) when
choosing $C_4 = 16 \mu^2$ and $C_5=4M$. This implies that the metric
function $f_-(r)$ becomes  the KS solution  $f_{\rm KS}(r)$ in
Eq.(\ref{sph_sol}) for slowly rotating black hole solution.
Therefore, our slowly rotating black hole solution is given by
\begin{equation}
ds^2_{\rm slow~KS} = -f_{\rm KS}(r) dt^2 + \frac{dr^2}{f_{\rm
KS}(r)} + r^2 d \theta^2 +
       r^2 \sin^2 \theta \left ( d \phi^2 - \frac{4J}{r^3} dt d\phi \right ).
\label{slowrsol}
\end{equation}
This is our main result.

 On the other hand,  the Kerr black hole is
given by
\begin{equation}
ds^2_{\rm Kerr} = - \frac{\rho^2 \Delta_r}{\Sigma^2} dt^2 +
\frac{\rho^2}{\Delta_r} dr^2
     + \rho^2 d\theta^2 + \frac{\Sigma^2 \sin^2 \theta}{\rho^2}
       \left ( d\phi - \xi dt \right )^2,
\label{kerr_metric}
\end{equation}
where
\begin{eqnarray}
\rho^2 &=& r^2 + a^2 \cos^2 \theta, \nonumber \\
\Delta_r &=& \left ( r^2 + a^2 \right ) - 2 M r, \nonumber \\
\Sigma^2 &=& \left ( r^2 + a^2 \right )^2 - a^2 \sin^2 \theta \Delta_r, \nonumber \\
\xi &=& \frac{2Mar}{\Sigma^2}.
\end{eqnarray}
In the slowly rotating limit of $J \ll M(a \ll 1)$, the Kerr
solution reduces to \be \label{SK} ds^2_{\rm
slow~Kerr}=-\Big(1-\frac{2M}{r}\Big)dt^2+\frac{dr^2}{\Big(1-\frac{2M}{r}\Big)}
+r^2d\theta^2+r^2\sin^2\theta\Big(d\phi^2-\frac{4J}{r^3}dt
d\phi\Big)\ee whose  metric  is given by \begin{equation}
g_{\mu\nu}^{\rm slow~Kerr}=g_{\mu\nu}^{\rm Sch}-\frac{2 Ma}{r}
\sin^2\theta \delta^t_\mu \delta^\phi_\nu.\end{equation} This means
that the slowly rotating Kerr black hole could be interpreted as
arising from the breaking of spherical to axial symmetry.  It is
easily checked that in the limit of $\omega \to \infty$, the slowly
rotating solution (\ref{slowrsol}) leads to the slowly rotating Kerr
solution (\ref{SK}).

Since we have still the KS metric function $f_{\rm KS}(r)$ for
slowly rotating black hole, we use  the horizon mass $M_h$ and
temperature $T_H$ for the KS black hole as~\cite{AdmMyung,WJDC} \be
M_h(r_+,\omega)=\frac{r_+}{2}-\frac{3 \tan^{-1}(\sqrt{\omega}
r_+)}{4\sqrt{\omega}},~~T_H=\frac{2\omega r_+^2-1}{8\pi r_+(\omega
r_+^2+1)}, \ee where $r_+$ is the outer horizon as a root of $f_{\rm
KS}(r_+)=0$. Importantly, we have angular velocity defined by
\begin{equation}
\Omega =-\frac{g_{t\phi}}{g_{\phi\phi}}= aN^\phi = \frac{2J}{r^3},
\label{rot_vel}
\end{equation}
which is the same form as that of the slowly rotating Kerr black
hole. The angular velocity on the horizon  is given by
\begin{equation}
\Omega_h = \frac{2M_h a}{r_+^3} = a \Bigg[ \frac{1}{r_+^2} -\frac{3
\tan^{-1}(\sqrt{\omega} r_+)}{2\sqrt{\omega}r_+^3}\Bigg]
\label{rot_bh}
\end{equation}
which reduces, in the limit of $\omega \to \infty$,
 to  the angular velocity of the slowly rotating Kerr
black hole on the horizon  \begin{equation} \Omega_h^{\rm Kerr} =
\frac{a}{r_+^2}=\frac{2J}{r_+^3} \label{rot_bh_kerr}
\end{equation}
 with $M_{\rm Sch}=r_+/2$.
Finally, the angular momentum of slowly rotating black hole is
given by
\begin{equation}   J=aM_h=a\Bigg[\frac{r_+}{2}-\frac{3 \tan^{-1}(\sqrt{\omega}
r_+)}{4\sqrt{\omega}}\Bigg]. \end{equation} In the limit of $\omega
\to \infty$, it leads to  the angular momentum of slowly rotating
Kerr black hole
\begin{equation}   J \to J^{\rm slow~Kerr}=\frac{ar_+}{2}. \end{equation}
If one uses the mass parameter defined in
(\ref{sph_sol})~\cite{Myung} instead of the horizon mass $M_h$, it
takes the form
\begin{equation}
M(r_+,\omega)=\frac{r_+}{2}+\frac{1}{4\omega r_+}. \end{equation}
Here we observe an inequality
\begin{equation}
M_h <M_{\rm Sch}<M, \end{equation} but they become the same form  in
the limit of $\omega\to \infty$.

\section{Discussions}
It seems to be a formidable task to find a fully rotating black hole
in the HL gravity because  full equations  to be solved are very
complicated.  In this work,  we have found the slowly rotating black
hole solution based on the KS solution in the HL gravity.

First of all, we explain why the slowly rotating solution is
naturally  obtained for the HL gravity by examining the order of
rotation parameter $a$ in equations of motion. The axisymmetric
metric ansatz (\ref{axi_ansatz}) was implemented by one component
shift vector, $N^\phi$.  Hence, extrinsic curvature $K_{ij}$ has
off-diagonal components as shown in (\ref{K_ij}).  This implies that
equation (\ref{eom1-KS}) obtained from variation of lapse function
$N$ remains unchanged when  adding  a rotating parameter term   to
the spherically symmetric case. This is confirmed by showing that
$K=0$ and $K_{ij}=0$  for spherically symmetric case, while for
slowly rotating case, $K=0$ and $K_{ij} K^{ij} = {\cal{O}}(a^2)$.
Effectively, Eq.(\ref{eom1-KS}) is the same equation  for both two
cases.  A  shift vector $N^{\phi}$  could be determined by
Eq.(\ref{eom2-KS}).  We emphasize that other equations remain
unchanged at  linear order of $a$. It is clear that $E^{(1)}_{ij}=0$
for spherically symmetric case and $E^{(1)}_{ij}={\cal{O}}(a^2)$ for
slowly rotating case, which is effectively taken to be zero.
$E^{(2)}_{ij}=0$ for  both two cases. All other $E^{(r)}_{ij}$ for
$r = 3, \cdots, 6$ remain unchanged, since they contain  terms
derived from $g_{ij}$ which do not carry rotation parameter $a$, and
$C_{ij} = 0$ by rotation symmetry in 3D Euclidean space.

In summary, all equations of motion remain unchanged by introducing
a slowly rotating parameter $a$ except Eq. (\ref{eom2-KS}), which
was  zero for spherically symmetric case. The slowly rotating black
hole could be interpreted as arising from the breaking of spherical
to axial symmetry.

Finally, we propose  a general  axisymmetric  metric ansatz for a
slowly rotating black hole \begin{equation} ds^2 = -N(r)^2 dt^2 +
\frac{dr^2}{f(r)} + r^2 d \theta^2
       + r^2 \sin^2 \theta \Big[ d \phi^2 - 2a N^\phi(r) dt
       d\phi\Big]
\label{axi-metric-general}
\end{equation}
with the shift vector
\begin{equation}
N^\phi(r) = C_1 + C_2 \int^r \frac{N(r')}{r'^4 \sqrt{f(r')}} dr' .
\label{sol_phi_general}
\end{equation}
Hence,  for any spherically symmetric solution,  we always construct
a slowly rotating  solution  by introducing a non-zero shift
function determined by  Eq. (\ref{sol_phi_general}). Considering
(\ref{axi-metric-general}),  any spherically symmetric solutions may
be also candidates for constructing  slowly rotating black hole
solutions in the HL gravity.  However, it seems difficult to  seek
the slowly rotating black holes  of the LMP black hole
solutions~\cite{LMP} because these solutions include the $\lambda=1$
AdS black hole with double horizons, $\lambda=1/2$ and 9/25 Lifshitz
black holes with single horizon, and $\lambda \ge 3$
Reissner-Nordstr\"om-type black holes with double
horizons~\cite{lifMyung}.

As another example, we would like to mention that the KS black hole
could be also  the solution to modified $F(R)$ HL
gravity~\cite{CNOOT,CCNOOT,COT}. Especially, in order to find the KS
black hole solution, the non-projectable modified $F(R)$ HL action
in~\cite{COT}
\begin{eqnarray}
S_F&=&\frac{1}{\kappa^2}\int dt d^3 x \sqrt{g}N F(^4\tilde{R}), \\
^4\tilde{R}&=&=K_{ij}K^{ij}-\lambda K^2
+2\tilde{\mu}\nabla_\mu(n^\mu\nabla_\nu n^\nu-n^\nu\nabla_\nu
n^\mu)-{\cal L}_R(g_{ij})
\end{eqnarray}
could be transformed into the KS action~\cite{AdmMyung}
\begin{equation} \mu^4 \int dt d^3 x \sqrt{g}N \tilde{{\cal L}}_{V}=\mu^4 \int dt d^3x \sqrt{g}N\Bigg[R
-\frac{2}{\omega}\Big(R_{ij}R_{ij}-\frac{3}{8}R^2\Big)\Bigg],
\end{equation}  by adjusting the parameters $\alpha_i (i=1,\cdots, 8)$ in ${\cal
L}_R(g_{ij})$ of (2.55) and taking $F(^4\tilde{R})=$
$^4\tilde{R},~K_{ij}=\tilde{\mu}=0$, and $1/\kappa^2=\mu^4$. In this
case, we propose that our solution (\ref{slowrsol}) is also the
slowly rotating solution to $F(R)$ HL gravity.   It is interesting
to find a spherically symmetric solution for $F(^4\tilde{R})\not=$
$^4\tilde{R}$ case and its slowly rotating solution.

\section*{Acknowledgement}
 This work was   supported  by Basic
Science Research Program through the National Research  Foundation
(NRF) of Korea funded by the Ministry of Education, Science and
Technology (2009-0086861).


\end{document}